\documentclass[12pt]{JHEP3}

\newcommand{\N}{{\scriptscriptstyle N}}

\newcommand{\be}{\begin{equation} }
\newcommand{\ee}{\end{equation} }
\newcommand{\ba}{\begin{array}}
\newcommand{\ea}{\end{array}}

\newcommand{\bea}{\begin{eqnarray}}
\newcommand{\eea}{\end{eqnarray}}

\newcommand{\nn}{\nonumber}

\def\P{{\cal P}}
\def\R{{\cal R}}

\usepackage{graphicx}
\usepackage{amssymb}
\def\beq{\begin{equation}}
\def\eeq{\end{equation}}
\def\bea{\begin{eqnarray}}
\def\eea{\end{eqnarray}}
\def\bef{\begin{figure}}
\def\enf{\end{figure}}
\def\S{{\bf S}}
\def\C{{\bf C}}
\def\Z{{\bf Z}}
\def\R{{\bf R}}
\def\N{{\bf N}}

\def\P{{\bf P}}

\def\ba{\begin{array}}
\def\ea{\end{array}}
\def\bce{\begin{center}}
\def\ece{\end{center}}
\def\nn{\noindent}

\def\be{\beta}

\def\Tr{{\rm Tr}\,}
\def\nn{\nonumber}

\title{Topological B-model and ${\hat c}=1$ String Theory}

\author{
Seungjoon Hyun$^{a}$\footnote{\tt hyun@phya.yonsei.ac.kr}, Kyungho
Oh$^{b}$\footnote{\tt oh@arch.umsl.edu}, Jong-Dae
Park$^{a}$\footnote{\tt jdpark@phya.yonsei.ac.kr} and Sang-Heon
Yi$^{c}$\footnote{\tt shyi@kias.re.kr}\\ \vskip1cm
{\it ${}^{a}$Institute of Physics and Applied Physics, Yonsei University, Seoul 120-749, Korea\\
${}^{b}$ Department of Mathematics and Department of Physics and
Astronomy, \\University of
Missouri-St. Louis, Saint Louis, MO 63121, USA\\
${}^{c}$Korea Institute for Advanced Study, Cheongnyangni 2-dong,
Dongdaemun-gu, Seoul, 130-722, Korea}}

\abstract{We study the topological B-model on a deformed $\Z_2$
orbifolded conifold by investigating variation of complex
structures via quantum Kodaira-Spencer theories. The
fermionic/brane formulation together with systematic utilization
of symmetries of the geometry gives rise to a free fermion
realization of the amplitudes. We derive Ward identities which
solve the perturbed free energy exactly. We also obtain the
corresponding Kontsevich-like matrix model. All these confirm the
recent conjecture on the connection of the theory with  ${\hat
c}=1$ type 0A string theory compactified at the radius
$R=\sqrt{\alpha'/2}$.}

\keywords{Topological B-Model, Deformed Orbifolded Conifold, 0A
String
  Theory}

\preprint{hep-th/0502075\\KIAS-P05018}

\begin{document}


\section{ Introduction}
Topological string theories on Calabi-Yau(CY) threefolds were
introduced~\cite{Witten:1988xj} to probe the topological nature of
the space. There are two different types, the A-model and the
B-model, of topological string theories, which study the
K\"{a}hler structure moduli space and the complex structure moduli
space of the CY threefolds, respectively. These theories have
proven to be very useful to study various aspects of critical and
noncritical string theories. Among others, they were used to compute
F-terms in superstring compactifications to four dimensions and
were shown to be equivalent to $c=1$ non-critical strings.
Furthermore, it was
shown~\cite{Dijkgraaf:2002fc,Dijkgraaf:2002vw,Dijkgraaf:2002dh}
that the topological B-model on local CY threefolds is large $N$
dual to the Dijkgraaf-Vafa(DV) matrix model, from which the
computation of nonperturbative superpotentials of ${\cal N}=1$
gauge theories reduces to perturbative computations in a DV matrix
model. For the recent review on the topological string theory,
see~\cite{Marino:2004uf,Marino:2004eq,Neitzke:2004ni}.

Recently, an extremely powerful method to solve exactly the
topological B-model on an interesting class of local CY threefolds
was developed in~\cite{Aganagic:2003qj}. It was shown that the
model is governed by the ${\cal W}$-algebra symmetries of the CY
threefolds, namely the holomorphic diffeomorphism. In particular,
it was powerful enough to show the full equivalence between the
topological B-model on deformed conifold and the $c=1$ noncritical
bosonic string theory at the self-dual radius. Both theories admit
the same free fermion description and have the symmetries which
characterize the theories completely with the emergence of dual
Kontsevich-like matrix model.

In somewhat different context, there have been interesting
developments~\cite{Takayanagi:2003sm,Douglas:2003up} in ${\hat
c}=1$ type 0 noncritical string theory. There are two different
types, so-called 0A and 0B, of noncritical string theories, both
of which have a matrix model description.  Since all the matrix
models are believed to be equivalent to the topological B-models
on some CY threefolds, it is natural to expect that there is a
topological B-model on some CY threefold which is equivalent to
${\hat c}=1$ type 0 string theory.  Indeed, it was suggested in an
interesting paper~\cite{Ita:2004yn} that ${\hat c}=1$ type 0A
string theory at the radius $R= \sqrt{\alpha'/2}$ is equivalent to
the topological B-model on the (deformed) $\Z_2$ quotient of the
conifold. In ~\cite{Ita:2004yn}  the ground ring structure of
${\hat c}=1$ string theory was identified with certain orbifolded
conifold. The partition functions, integrable structures and the
associated Ward identities of ${\hat c}=1$ theory were considered
and the corresponding Kontsevich-like matrix model was
constructed. From the topological B-model on the $\Z_2$ orbifolded
conifold, several results are obtained leading to the stated
conjecture. However, because the $\Z_2$ orbifolded conifold has
non-isolated singularities, topological strings are not
well-defined and thus the computation in ~\cite{Ita:2004yn}
is, at best, suggestive.

In this paper, we consider the topological B-model on the deformation
of $\Z_2$ orbifolded conifold, which is well-defined since the deformed geometry is smooth. We adopt the method given
in~\cite{Aganagic:2003qj} to study the topological B-model and
confirm that the model is indeed equivalent to the ${\hat c}=1$
type 0A string theory compactified at the radius  $R=
\sqrt{\alpha'/2}$ with non-vanishing RR flux background. The key
role is played by the ${\cal W}$-algebra symmetry which comes from
the holomorphic diffeomorphism of the CY threefolds. The
$\hat{c}=1$ type 0 string theory also enjoys the same kind of
${\cal W}$-algebra symmetry.  As this symmetry is powerful enough
to constrain the whole theory, we can regard this, more or less,
as a proof of their equivalence. Indeed we will show that the
topological B-model on that space has the free  fermion
description just like the ${\hat c}=1$ theory. We use the symmetry
to derive the Ward identities from which we determine the
perturbed partition function of the topological B-model under complex
deformations. It matches exactly with the generating functional of
${\hat c}=1$ matrix model under the perturbation by tachyon
momentum modes. This also guarantees that both theories admit the
same Kontsevich-like matrix model description.

As will be clear, the RR flux in ${\hat c}=1$ 0A string theory
corresponds to a deformation parameter of the $\Z_2$ orbifolded
conifold which makes the space non-singular. Therefore the
non-vanishing RR flux is crucial in order to have a well-defined
topological B-model on the non-singular space. We can consider the
limit where the corresponding deformation parameter vanishes. The
resultant formulae we obtain in this paper are well-defined under
the limit and correspond to the formulae in ${\hat c}=1$ theory
with vanishing RR flux.

The organization of this paper is as follows: In section 2, we
explain the prescription to solve the topological B-model on some
class of local CY threefolds. In particular, we take an example of
the model on the deformed conifold which is equivalent to the $c=1$
bosonic string theory compactified at the self-dual radius. Using
this example we explain salient features of those models which
show their equivalence most clearly. In section 3, we turn to
${\hat c}=1$ 0A matrix model and describe the relevant features
for our study on the topological B-model. In section 4, firstly we
review the deformation and the resolution of the orbifolded
conifold, which give rise to non-singular geometries on which
string theory can be well-defined. And then we explain the basic
set-up to solve the topological B-model on the deformation of
$\Z_2$ orbifolded conifold. Using this, in section 5, we study the
integrable structure of the model and show that it is exactly the
same as the one of the ${\hat c}=1$ 0A string theory. We show that
both theories have the same Ward identities and thus the same
perturbed free energy and, as a result, correspond to the same
Kontsevich-like matrix model. In section 6, we draw our
conclusions and further comments.

\section{Topological B-model on the Local Calabi-Yau
threefold}\label{sec2}

In this section we review the topological B-model on non-compact
CY geometries and explain general strategies to solve the model on
some class of CY geometries following~\cite{Aganagic:2003qj}. In
particular we focus on the topological B-model on the deformed
conifold which was shown to be equivalent to $c=1$ bosonic string
theory compactified at the self-dual radius. These models share a lot
of characteristic features with those models we consider in this paper.

\subsection{Basic set-up}\label{sec2.1}
The topological B-model describes the quantum theory of the
complex structure deformation  of Calabi-Yau geometries, which
corresponds to the quantum Kodaira-Spencer theory of
gravity~\cite{Bershadsky:1993cx}. In the theory we consider maps
from a Riemann surface $\Sigma_g$ of genus $g$ to the target CY
manifold. For each Riemann surface $\Sigma_g$, we compute the
corresponding free energy, ${\cal F}_g(t_i)$,  where $t_i$'s denote the
complex structure deformation parameters. Those free energies are
summed with weight $g_s^{2g-2}$ to give the free energy of the
B-model topological closed strings for all genus as
\begin{equation}
{\cal F}(g_s,t_i) = \sum_{g=0}^{\infty} g_s^{2g-2}{\cal F}_g(t_i)\,,
\end{equation}
where $g_s$ is the string coupling constant.

Recently, the B-model topological string theory on the local CY
threefold of the form
\begin{equation} \label{hyper1}
uv- H(x, y)=0
\end{equation}
was studied by considering ${\cal W}$-algebra symmetries underlying in
this type of CY threefolds~\cite{Aganagic:2003qj}.
These symmetries correspond to the holomorphic diffeomorphisms of the
target CY threefold which preserve the  equation of CY threefold
(\ref{hyper1}) and the holomorphic three-form\footnote{The
  normalization of the holomorphic three-form is chosen for later
  convenience.}
\begin{eqnarray}\label{3-form}
\Omega =\frac{1}{4\pi^2}\frac{dx\wedge dy \wedge du}{u}~.
\end{eqnarray}
The CY geometry (\ref{hyper1}) can be viewed as a fibration over
the ($x, y$) plane with one dimensional fibers.  $H(x,y)=0$ in the
base manifold is the locus where  the fiber degenerates into two
components $u=0$ and $v=0$. By integrating along a contour around
$u=0$, the periods of the three-form $\Omega$ over three-cycles
$C_i$ become integrals of the two-form \bea \int_{D_i} dx \wedge
dy \eea over domains $D_i$ in the $(x,y)$-plane whose boundary is
one-cycles $c_i$ on the algebraic curve $H(x,y) =0$. Hence by
Stokes' theorem, the periods reduce to integrals of the one-form
\begin{equation}
\int_{c_i} y dx~.
\end{equation}
 Therefore the CY geometry is characterized by the
algebraic curve $H(x,y)=0$ and thus the complex deformations of the CY geometry
are captured by the canonical one-form, $ydx$.

If non-compact B-branes wrap the fiber, the worldvolume theory of
the B-branes is  given by a dimensional reduction of holomorphic
Chern-Simons theory to one complex
dimension~\cite{Aganagic:2000gs}. It becomes the theory of the
Higgs fields $x(u)$ and $y(u)$ whose zero modes are identified
with coordinates of the moduli space of these non-compact
B-branes, which is nothing but the base manifold. In the action,
the  Higgs fields $x(u)$ and $y(u)$ play roles as canonically
conjugated fields and thus their zero modes have a canonical
commutation relation
\begin{equation}
[x,y]= i g_s\,. \label{ComRel}
\end{equation}

All these imply that $x$ and $y$ are conjugate variables which
form a symplectic pair with symplectic structure $dx \wedge dy$.
After the reduction to the base manifold, ($x, y$) plane, the
holomorphic diffeomorphisms that preserve the equation
(\ref{hyper1}) of the CY geometry and the corresponding
holomorphic three form $\Omega$ descend to the diffeomorphisms of
the plane that preserve the symplectic form $dx\wedge dy$, and
therefore are given by general holomorphic canonical
transformations on ($ x, y$).

The complex structure deformation of the curve $H(x,y)=0$ can
appear only at `infinity' i.e. only at the boundary if the
`compactified' curve is a Riemann surface of genus zero which does
not have any complex deformation moduli.  Let us introduce a local
coordinate $x$ near each boundary such that $x\rightarrow \infty$
at the boundary. Near each boundary the canonical one-form is
given by $ydx$ which can be identified with the Kodaira-Spencer
field $\phi(x)$ as
\begin{equation}
ydx = \partial \phi~.
\end{equation}
The background field $\phi_{cl}(x)= \langle\phi(x)\rangle$ can be
determined from the relation $H(y_{cl}(x),x)=0$ as $y_{cl}(x)=
\partial_x\phi_{cl}$. Arbitrary deformations of chiral bosonic
scalar field $\phi(x)$ which correspond to complex deformations of
the curve near the boundaries are given by the reidentification of
$y(x)$ or, equivalently, by mode expansion of
$\phi(x)$ around $x\rightarrow \infty$ of the form
\begin{eqnarray}
y(x)=\partial_x\phi(x)=y_{cl}(x) + t_0 x^{-1} -
g_s\sum_{n=1}^{\infty}n t_nx^{n-1} +g_s
\sum_{n=1}^{\infty}\frac{\partial}{\partial
  t_n}
x^{-n-1}~.
\end{eqnarray}

The quantum free energy ${\cal F}$ is a function of the infinite
set of couplings $t_n^i$, where $i$ labels the boundaries. Then
one can regard the quantum free energy ${\cal F}$ as a state
$|V\rangle$ in the Hilbert space ${\cal H}^{\otimes k}$, where
${\cal H}$ is the Hilbert space of a single free boson and $k$ is
the number of boundaries~\cite{Witten:1993ed}. One convenient
representation of the state $|V\rangle$ is a coherent state
representation. In terms of the standard mode expansion of a
chiral boson
\begin{eqnarray}
\partial_x\phi(x)=-i\sum_{n\in {\bf Z}} \alpha_nx^{-n-1}~, \qquad
[\alpha_n, \alpha_m] =n g_s^2\delta_{n+m, 0}~,
\end{eqnarray}
the coherent state $|t\rangle$ is defined as
\bea
 |t\rangle \equiv
\exp\Big[\frac{i}{g_s}\sum_{n=1}^{\infty}t_n\alpha_{-n}\Big]|0\rangle~.
\eea
In this coherent state representation, the partition function can be
expressed as
\bea
Z(t^i) = \exp {\cal F}(t^i) = \langle t^1|\otimes\cdots\otimes\langle t^k|V\rangle~,
\eea
where $k$ is the number of boundaries.

One way to perturb the geometry is to insert B-branes near the
boundaries. Non-compact B-branes which wrap the fiber and reside on the locus
$H(x, y)=0$  can be realized in the closed string sector by
introducing
(anti-)brane creation operator, $\psi(x)$ ($\psi^*(x)$):
\begin{equation}\label{B-brane}
\psi(x) =~ : \exp \Big[-\frac{i}{g_s}\,\phi(x)\Big] :~, \qquad
\psi^*(x) =~ : \exp \Big[\frac{i}{g_s}\,\phi(x)\Big] :~,
\end{equation}
which is the fermionization of Kodaira-Spencer field $\phi$~\cite{Aganagic:2003db}.

Let us consider two patches with symplectic pairs of coordinates ($x_i, y_i$)
and ($x_j, y_j$), respectively,  and the corresponding fermions. Those
two symplectic pairs are related by a canonical transformation
preserving the symplectic form $dx_i\wedge dy_i = dx_j\wedge dy_j$
with a generating function $S(x_i, x_j)$. The corresponding fermions
transform just like the wave function on the geometry, and hence the
transformation of fermions between two patches, $i$, $j$ is given by~\cite{Aganagic:2003qj}
\begin{equation}\label{trans1}
\psi_{j}(x_j) = \int dx_i~
e^{\frac{i}{g_s}\,S(x_i,x_j)}\psi_i(x_i)\,.
\end{equation}
In the next subsections we consider an explicit example in which the
quantum free energy of the topological B-model can be computed
explicitly in this setup.

\subsection{The $c=1$ bosonic string}\label{sec2.2}

In this subsection we review some aspects of the matrix model
description of $c=1$ bosonic string theory which are relevant for
our study. For the review on various aspects of $c=1$ theory,
see~\cite{Klebanov:1991qa,Ginsparg:1993is,DiFrancesco:1993nw,
Polchinski:1994mb,Alexandrov:2003ut}.

The $c=1$ bosonic string
theory compactified at the self-dual radius is
equivalent~\cite{Witten:1991zd,Ghoshal:1995wm,Dijkgraaf:2003xk} to the
topological B-model on the deformed conifold which is given by the
hypersurface (\ref{hyper1}) with
\begin{equation}
H(x, y)= xy-\mu~,
\end{equation}
where $\mu$ is the complex deformation parameter. 
The first indication of this equivalence came from the nonchiral
ground ring of  $c=1$ bosonic string at the self-dual
radius~\cite{Witten:1991zd}. The
operator product of the spin zero ghost number zero BRST invariant
operators is again BRST invariant and gives a commutative and
associative ring structure, modulo BRST exact terms, as
\begin{equation}
{\cal O}(z) {\cal O}^\prime (0) \sim {\cal O}^{\prime \prime} (0) + \{Q, ...\}~.
\end{equation}
This is called the ground ring.  The right- and left-moving sectors of
$c=1$ bosonic string give chiral rings and
thus, in combination, the corresponding closed string theory has
nonchiral ground ring.  At the self-dual radius, it is
generated by
four operators $u,v,x,y$, which correspond to tachyon momentum and
winding states, and they obey the relation
\begin{equation}
uv-xy=\mu_{M}~,
\end{equation}
where $\mu_{M}$ is the level of the Fermi sea of the $c=1$ matrix
model.

Noncritical $c=1$ bosonic string theory has a free-fermion
description with inverted harmonic oscillator potential. The
Hamiltonian is given by
\begin{eqnarray}
H =  \frac{1}{2} (p^2- x^2)~,
\end{eqnarray}
in the usual $\alpha'=1$ convention of $c=1$ bosonic string theory. It
is convenient to introduce the
light-cone variables $x_{\pm}= (x\pm p)/\sqrt{2}$, in terms of
which the Hamiltonian
becomes~\cite{Alexandrov:2002fh,Alexandrov:2003qk,
Alexandrov:2003nn,Alexandrov:2004cg}
\beq H = -\frac{1}{2}(x_+x_-+x_-x_+)\,.  \eeq
In this formulation it is clear that $x_-$ and $x_+$ are conjugate
with commutation relation, $ [x_-,x_+]=i$ and thus the
Schr\"{o}dinger wave function can be represented by either $x_+$
or $x_-$. Let us denote the wave function in the $x_+$ and $x_-$
representations as $\psi_+$ and $\psi_-$, respectively. The energy
eigenfunctions are given by
\beq \psi_{\pm}^E(x_\pm) = \frac{1}{\sqrt{2\pi}}\, x_{\pm}^{\pm iE-1/2}\,.
\eeq
The vacuum state of the system corresponds to the Fermi sea in
which fermions are filled in the left-hand side of inverted
harmonic oscillator. The correlation functions of tachyon
operators in $c=1$ string theory correspond to the perturbation of the
Fermi sea in the matrix quantum mechanics.

In the $\cal{S}$-matrix formulation of the $c=1$ matrix model,
right-moving modes,  $\psi_-(x_-)$,  and
left-moving modes, $\psi_+(x_+)$,
correspond to incoming and outgoing excitations,
respectively. These in and
out wave functions are related by ${\cal S}$-matrix as
\beq \psi_{+}(x_+) = (S\psi_{-})(x_+)= \int dx_-~ K(x_+,x_-)
\psi_{-}(x_-)\,, \label{InOut}\eeq
where the kernel, $K(x_-,x_+)$ can be taken as
$e^{ix_+x_-}/\sqrt{2\pi}$. Note that one may choose different
kernel such as  $\sqrt{\frac{2}{\pi}}\cos(x_+x_-)$ or $i
\sqrt{\frac{2}{\pi}}\sin(x_+x_-)$. All these kernels give the same
results modulo non-perturbative terms. Since  we are concerned
about the free energy only at the perturbative level, we can
choose any one of them as a kernel. It is convenient to use the exponential
kernel for matching the corresponding expression from the B-model
topological string side. Also note that, since fermions fill in
the left-hand side of the inverted harmonic oscillator potential,
it is reasonable to take the integration region along the negative
real axis\footnote{When fermions fill the right-hand side of
inverted harmonic oscillator, the positive real axis is taken as
the integration
region~\cite{Alexandrov:2003ut,Alexandrov:2002fh}.}. Reflection
coefficient can be introduced as
\beq {\cal R}(E)\psi_{+}^E(x_+) =
(S\psi_{-}^E)(x_+)=\frac{1}{\sqrt{2\pi}}\int_{-\infty}^0 dx_-~
e^{ix_+x_-} \psi_{-}^E(x_-)\,, \eeq
{}from which it can be computed as
\begin{equation}
{\cal R}(E)  =-\frac{1}{\sqrt{2\pi}}\, e^{i\frac{\pi}{2}
(-iE+\frac{1}{2})} \Gamma(-iE +\frac{1}{2})\,. 
\end{equation}

One can euclideanize the time coordinate in the target space and consider the
compactification of the theory with compactification radius $R$.
The above reflection coefficient can be used to obtain the `free energy' of
the model.
The perturbative `free energy' of grand canonical ensemble is defined by
\begin{equation}
{\cal F}(\mu) = \int^{\infty}_{-\infty} d E~ \rho (E) \ln
[1+e^{-2\pi R(E+\mu)}]\,,
\end{equation}
where $\rho(E)$ is the density of states and $\mu$ is a chemical
potential. The relation between the density of states and the
reflection coefficient is known to
be~\cite{Alexandrov:2003qk,Douglas:2003up,Alexandrov:2003ut}
\begin{equation}
\rho (E) = \frac{1}{2\pi} \Big[ \frac{d \phi (E)}{d E} -\ln \Lambda\Big]\,,
\end{equation}
where $\phi(E)=  Im\ln {\cal R}(E)$ and $\Lambda$ is a cut-off.

In order to get an expression free from cut-off dependence, it is
convenient to consider the third order derivative of the free energy
\bea  \frac{\partial^3{\cal F}}{\partial \mu^3} &=& -2\pi R
\int^{\infty}_{-\infty} d E~ \frac{d^2 \rho(E)}{d
E^2}\frac{1}{e^{2\pi R(E+\mu)}+1}\,. \eea
In the $c=1$ matrix model, we have
\beq \frac{d^3\phi_{c=1}(E)}{d E^3} = {\rm I m}\Big[(-i)^3
\psi^{(2)}\Big(-i E+\frac{1}{2}\Big)\Big]\,, \eeq
where the polygamma function $\psi^{(n)}$ is defined as
\beq \psi^{(n)}(z) \equiv \frac{d^{n+1}}{d z^{n+1}} \ln \Gamma(z)=
(-1)^{n+1}\int^{\infty}_{0} dt~ \frac{t^n e^{-zt}}{1-e^{-t}}\,.
\eeq
The computation of $ \frac{\partial^3}{\partial \mu^3_M}{\cal
F}_{c=1}(\mu_M) $ is straightforward, given by the contour
integral in the upper half-plane, and leads to
\bea  \frac{\partial^3{\cal F}_{c=1}(\mu_M)}{\partial \mu^3_M} &=&
-2\pi R \int^{\infty}_{-\infty} d E~ \frac{d^2 \rho_{c=1}(E)}{d
E^2}\frac{1}{e^{2\pi R(E+\mu_M)}+1} \nn \\
&=& R~ {\rm I m }\bigg[ \int^{\infty}_{0}d t~ e^{-i\mu_M t}\,
\frac{t/2}{\sinh (t/2)}\frac{t/(2R)}{\sinh (t/2R)}\bigg]\,. \eea
The resultant perturbative free energy of $c=1$ matrix model at
the self-dual radius is given by
\begin{eqnarray}
{\cal F}_{c=1} (\mu_M) = -\frac{1}{2}\mu_M^2 \ln \mu_M -\frac{1}{12} \ln \mu_M
+\sum_{g\geq 2}^{\infty}\frac{|B_{2g}|}{2g(2g-2)} \mu_M^{2-2g}~,
\end{eqnarray}
where
\begin{equation}
B_{2g} = (-1)^{g-1}|B_{2g}| = \frac{(-1)^{g-1} 2 (2n)!}{(2\pi)^{2n}}\zeta (2n)~.
\end{equation}

\subsection{The Topological B-model on the deformed conifold}\label{sec2.3}
As explained in sect.~\ref{sec2.1}, the Kodaira-Spencer theory on the
deformed conifold which corresponds to $c=1$ strings at the
self-dual radius is described by the chiral boson on the Riemann
surface, \bea \label{curve1} H(x, y) =xy-\mu=0~. \eea The period
integrals over the symplectic basis of the three-cycles in the
case of CY geometry of the type (\ref{hyper1}) are given by
\begin{eqnarray}
X^i =\oint_{A^i} \Omega = \frac{1}{2\pi i} \oint_{a^i} ydx\,,
\qquad F_i = 2\pi i \int_{B^i} \Omega =  \int_{b^i} y dx\,.
\end{eqnarray}
These pairs of periods are related by the following relation
\begin{equation}
F_i = \frac{\partial {\cal F}_0}{\partial X^i}~,
\end{equation}
where $ {\cal F}_0$ denotes the tree level free energy or prepotential.

We need to introduce a cut-off $\Lambda$ for the integral over
noncompact $B$-cycle ($b$-cycle). Then the period, for the curve (\ref{curve1}),
 can be computed as
\begin{eqnarray}
X = \frac{1}{2\pi i} \oint_{a} ydx = \mu\,, \qquad F =
2\int_{\sqrt{\mu}}^{\sqrt{\Lambda}} y dx =\mu \ln
\Big(\frac{\Lambda}{\mu}\Big)~,
\end{eqnarray}
from which one can easily read off the
tree level free energy as
\begin{eqnarray}
{\cal F}_0 =\frac{1}{2} X F= -\frac{1}{2} \mu^2\ln \mu +{\cal O} (\Lambda)~.
\end{eqnarray}

The full free energy of the topological B-model, all order in the
string coupling, can be obtained via Dijkgraaf-Vafa(DV) conjecture
on the equivalence between the topological B-model and the DV matrix
model~\cite{Dijkgraaf:2002fc,Dijkgraaf:2002vw,Dijkgraaf:2002dh}.
The DV matrix model dual to the topological B-model on the deformed
conifold is known to be the Gaussian matrix model whose partition
function is given by
\begin{eqnarray}\label{DV}
Z = e^{\cal F} = \frac{1}{{\rm Vol}( U(N))} \int {\cal D} M \ \
e^{-\frac{1}{2g_s} \Tr M^2}~.
\end{eqnarray}
In this correspondence, the 't Hooft coupling $t=g_s N$ is identified
with $i\mu$ in the topological B-model.
In the 't Hooft limit, the free energy can be explicitly computed and is given by
\begin{eqnarray}
{\cal F}_{DV}(t) = \frac{1}{2} \Big(\frac{t}{g_s}\Big)^2 \Big(\ln t-
  \frac{3}{2}\Big) -\frac{1}{12} \ln t + \sum_{g\geq
  2}^{\infty}\frac{B_{2g}}{2g(2g-2)} \Big(\frac{t}{g_s}\Big)^{2-2g}~.
\end{eqnarray}
Therefore we have
\begin{equation}\label{DV1}
{\cal F}_{top} (\mu=g_s\mu_M) = {\cal F}_{DV} (t=ig_s\mu_M) = {\cal F}_{c=1} (\mu_M)~,
\end{equation}
up to irrelevant regular terms.

 The Riemann surface (\ref{curve1}) has two boundaries and  can
be described by two patches whose coordinate is chosen to be $x$ and $y$,
respectively. Those two asymptotic regions, which correspond to
$x\rightarrow \infty$ and $y\rightarrow \infty$, may describe
the incoming and outgoing states in the $c=1$ matrix model,
respectively.   As
explained in the previous subsection, we introduce Kodaira-Spencer
fields in each patch as
\beq y=\partial_{x}\phi(x)\,, \qquad x =
-\partial_{y}\tilde\phi(y)\,. \eeq
Therefore, from the correspondence with $c=1$ matrix model,
$\partial_x\phi(x)$ can be regarded as ``incoming'' modes and
$\partial_y\tilde{\phi}(y)$ as ``outgoing'' ones.

The classical part of $\phi$ which describes the original
background geometry is given by
\begin{eqnarray}
\partial_{x}\phi_{cl}(x)=\frac{\mu}{x}\,,
\qquad  \partial_{y}\tilde{\phi}_{cl}(y) =
-\frac{\mu}{y}\,.
\end{eqnarray}
The quantum parts of $\phi$ describe the complex deformations of
the geometry  and have mode expansions
\begin{eqnarray}
\phi_{qu}(x) &=& -g_s\sum_{n=1}^{\infty}t_n x^n -g_s
\sum_{n=1}^{\infty}\frac{1}{n}\frac{\partial}{\partial t_n}
x^{-n}\,, \nonumber \\
\tilde{\phi}_{qu}(y)&=& -g_s\sum_{n=1}^{\infty}\tilde{t}_ny^n -
g_s \sum_{n=1}^{\infty}\frac{1}{n}\frac{\partial}{\partial
\tilde{t}_n}
y^{-n}\,,
\end{eqnarray}
where, in the classical limit, the couplings are given by periods
\bea g_s t_n = -\frac{1}{n} \oint_{x\rightarrow \infty}
 y x^{-n}~, \qquad
g_s \frac{\partial {\cal F}_0}{\partial t_n} = \oint_{x\rightarrow
\infty}
 yx^{n}~.
\eea

Non-compact B-brane
creation operator in each patch can be introduced
as (\ref{B-brane}). Suppose we put branes at positions $y= y_i$ near the
boundary $y\rightarrow \infty$, then the gravitational backreaction is
given by
\begin{eqnarray}\label{prod2}
\prod_{i=1}^N{\tilde \psi}(y_i) = \prod_i :\exp \Big[-\frac{i}{g_s}\,{\tilde \phi}(y_i)\Big]:
= \Delta(y) :\exp \Big[-\frac{i}{g_s}\sum_{i=1}^N{\tilde \phi}(y_i)\Big]:~,
\end{eqnarray}
where $\Delta (y)=\prod_{i<j}(y_i-y_j)$ denotes the Vandermonde
determinant.
The expectation value of $\partial {\tilde \phi}(y)$ in this perturbed
background becomes
\begin{eqnarray}
\langle \prod_{i=1}^N \tilde{\psi}(y_i)\partial\tilde{\phi}(y)\rangle
= -ig_s \sum_{i=1}^N \frac{1}{y_i-y} =
-ig_s\sum_{i=1}^N\sum_{n=1}^\infty y_i^{-n}y^{n-1}~.
\end{eqnarray}
This tells us that we can use B-branes to deform the curve with the coupling
\begin{eqnarray}\label{coeff1}
\tilde{t}_n = \frac{i}{n}\sum_{i=1}^N y_i^{-n}~.
\end{eqnarray}

Since $x$ and $y$ are conjugate each other as shown in
(\ref{ComRel}), they play dual roles as a coordinate in one patch and
a momentum in the other, and vice versa. Therefore the canonical
transformation between two coordinate patches, which leads to the
transformation (\ref{trans1}), is nothing but the Fourier transform:
\beq \tilde{\psi}(y) = (S\psi)(y) = \frac{1}{\sqrt{2\pi}}\int dx~
 e^{iyx/g_s}\psi(x)\,, \eeq
which is reminiscent of $c=1$ relation in ~(\ref{InOut}).

One can compute the quantum free energy of the topological B-model on the
deformed conifold, or the corresponding state $|V\rangle$, using these
B-branes and ${\cal W}$ symmetry of the
model. Ward identities associated with ${\cal W}$ symmetry are
enough to fix the quantum free energy. The curve $xy=\mu$ has two
punctures and, in this case, the ${\cal W}$ symmetry generators relate
an operation on one puncture to the one on the other. For example, consider an
action of the ${\cal W}_{1+\infty}$ generator given by the Hamiltonian
\begin{equation}
f(x_i, p_j) = x_i^n~,
\end{equation}
which can be written in the fermionic representation as
\begin{equation}
W_m^{n+1}= \oint \psi (x_i)x_i^n \psi^*(x_i)~.
\end{equation}
The corresponding Ward identity is given by~\cite{Aganagic:2003qj}
\begin{equation}
\oint_{x\rightarrow\infty} dx~ \psi(x)x^n \psi^*(x)|V\rangle =
-\oint_{y\rightarrow\infty} dy~
\tilde{\psi}(y)(ig_s\partial_{y})^n\tilde{\psi}^*(y) |V\rangle\,.
\end{equation}
This is identical with the Ward identity~\cite{Dijkgraaf:1992hk}
of the $c=1$ string amplitude. This ensures that both theories are
equivalent, sharing the same integrable structure.

After successive integrations by parts, the Ward
identity can have the following alternative form:
\beq \oint_{x\rightarrow\infty} dx~ \psi^*(x)x^n \psi(x)|V\rangle
= -\oint_{y\rightarrow\infty} dy~
\tilde{\psi}^*(y)(-ig_s\partial_{y})^n\tilde{\psi}(y) |V\rangle\,.
\eeq
In the following sections, we will use this form of Ward
identity to show the equivalence between
the topological B-model on the deformed $\Z_2$ orbifolded conifold and
the compactified ${\hat
  c}=1$ 0A string theory.

\section{The ${\hat c}=1$ type 0A string theory}\label{sec3}

The ${\hat c}=1$ string has ${\cal N}=1$ superconformal symmetry
on the worldsheet with one scalar superfield $X$, whose bosonic
component, $x$, corresponds to the time coordinate in the target
manifold, and one super Liouville field $\Phi$, which comes from
the worldsheet supergraviton multiplet. Under the nonchiral GSO
projection, it becomes either type 0A or 0B
theory~\cite{Takayanagi:2003sm,Douglas:2003up}. The spectrum of
0A(0B) theory consists of massless tachyon field and R-R
vector(scalar) fields. One can euclideanize the time coordinate,
$x$, of the target manifold and consider the circle
compactification with compactification radius $R$. After the
compactification, these two theories, type 0A and type 0B theories
are T-dual under $R\rightarrow \frac{\alpha'}{R}$.

In this paper we focus on the type 0A theory, especially at the
radius $R=\sqrt{\alpha'/2}$, which must be equivalent to type 0B
theory at the dual radius $R= \sqrt{2\alpha'}$. After the
compactification, the NS-NS spectrum of type 0A theory consists of
tachyon field momentum states with momentum $k=n/R$ and winding
states with $k=wR/\alpha'$ where $n, w$ take integer values, while
in the R-R sector, the theory has winding modes with
$k=wR/\alpha'$, only~\cite{Douglas:2003up}.

The ground ring is
generated by four elements, $u$, $v$, $x$ and $y$, among which $u$
and $v$ come from NS-NS momentum modes and $x$ and $y$ from R-R
winding modes. It was argued in~\cite{Ita:2004yn} that the ground ring
of ${\hat c}=1$ type 0A string theory at the radius
$R=\sqrt{\alpha'/2}$ is given by
\begin{equation}
uv-(xy-\mu)^2-\frac{q^2}{4}=uv - (xy-\mu+\frac{i}{2}q)(xy-\mu-\frac{i}{2}q) =0~,
\end{equation}
where $\mu$ is the cosmological constant and $q$ is the net
D0-brane charge in the background. This is the same form as the
equation of the deformed $\Z_2$ orbifolded conifold which will be
studied in detail using the topological B-model.

In this section we review some aspects of the matrix model description
of the ${\hat c}=1$ 0A string theory, which will be relevant in connection
with the topological B-model we will consider.

\subsection{Type 0A matrix quantum mechanics}\label{sec3.1}
The matrix model description of the ${\hat c}=1$ 0A theory is
given by the world volume theory of $N+q$ D0-branes and $N$
anti-D0-branes, which is $U(N+q) \times U(N)$ matrix quantum
mechanics~\cite{Douglas:2003up}:
\begin{eqnarray}
L= \Tr \Big[(D_0 t)^\dagger D_0 t + \frac{1}{2\alpha'} t^\dagger
t\Big]~,
\end{eqnarray}
where $t$ denotes the tachyon field in the bifundamental representation
under $U(N+q) \times U(N)$. This model can be described by a
non-relativistic free-fermion in two dimensions with upside-down
harmonic oscillator potential. The single particle Hamiltonian is
given by
\begin{eqnarray}
H =  \frac{1}{2} (p_x^2+p_y^2)- \frac{1}{4\alpha'}(x^2+y^2)~.
\end{eqnarray}
The conserved charge of the angular momentum $J= xp_y-yp_x$ is
identified with the net D0-brane charge $q$~\cite{Douglas:2003up}.
Note that in each sector of the angular momentum $J=q$, the model
becomes~\cite{Douglas:2003up,Kapustin:2003hi,DeWolfe:2003qf}
effectively one-dimensional model which is known as deformed
matrix quantum mechanics~\cite{Jevicki:1993zg} with the
Hamiltonian,
\begin{eqnarray}
H'= -\frac{1}{2}\frac{d^2}{dr^2} - \frac{1}{4\alpha'}r^2 +
\frac{q^2-\frac{1}{4}}{2r^2}~,
\end{eqnarray}
where $r=\sqrt{x^2+y^2}$.

It is again convenient to introduce the light cone
variables~\cite{Yin:2003iv}
\begin{eqnarray}
z_{\pm} = \frac{1}{\sqrt{2}}
\Big[\frac{1}{\sqrt{2\alpha'}}(x+iy)\pm (p_x +i p_y)\Big]~,
\end{eqnarray}
and their complex conjugates ${\bar z}_{\pm}$. The only nontrivial commutators are
\begin{eqnarray}
[z_+, {\bar z}_- ] = [ {\bar z_+}, z_- ] =
-\frac{2i}{\sqrt{2\alpha'}}~,
\end{eqnarray}
which tells that ($z_+, {\bar z}_+$) and ($z_-, {\bar z}_-$) form
conjugate pairs. Therefore in terms of these new variables, the
wave function can be expressed either in $( z_+, {\bar z_+})$
representation or in $( z_-, {\bar z_-})$ representation, denoted
by $\psi_+(z_+, {\bar z}_+)$ and $\psi_-(z_-, {\bar z}_-)$,
respectively. Furthermore the wave functions in $( z_+, {\bar
z_+})$ representation and those in  $( z_-, {\bar z_-})$
representation should be related by the Fourier transform:
\begin{eqnarray}\label{Fourier2}
\psi_+(z_+,\bar{z}_+) =
\frac{1}{2\pi}\int^{\, 0}_{-\infty} dz_-d\bar{z}_-
e^{i\bar{z}_+z_- +i z_+\bar{z}_-}\psi_-(z_-
 ,\bar{z}_-)\,,
\end{eqnarray}
where the integration region is taken along the negative real axis,
in the same fashion as the $c=1$ case.

From now on we set $\alpha'=2$, which is the standard convention for
${\hat c}=1$ theory. In the $( z_+, {\bar z_+})$ representation,
the Hamiltonian and the angular momentum can be expressed as
\begin{eqnarray}\label{ham1}
H= \frac{i}{2} \Big(z_+ \frac{\partial}{\partial z_+} + {\bar
z}_+\frac{\partial}{\partial {\bar z}_+} + 1\Big)~,
\end{eqnarray}
and
\begin{eqnarray}\label{ang1}
J= z_+ \frac{\partial}{\partial z_+} - {\bar z}_+\frac{\partial}{\partial {\bar z}_+}~.
\end{eqnarray}
Therefore the wave functions in the sector of angular momentum  $q$
are of the form
\begin{eqnarray}
\psi_{\pm} ( z_{\pm}, {\bar z}_\pm ) =
\Big(\frac{z_\pm}{\bar{z}_\pm}\Big)^{q/2} f(z_\pm \bar{z}_\pm )~.
\end{eqnarray}
The energy eigenstates with energy $E$ and angular momentum $q$
are given by
\begin{eqnarray}
\psi_{\pm}^{E, q} ( z_{\pm}, {\bar z}_\pm ) =
\Big(\frac{z_\pm}{\bar{z}_\pm}\Big)^{q/2} (z_\pm \bar{z}_\pm
)^{\pm iE
  -\frac{1}{2}}~.
\end{eqnarray}

As will be clear, this is the most natural approach when we compare
this theory with
the topological B-model on the deformed $\Z_2$
orbifolded conifold. As was the case in the $c=1$ matrix quantum
mechanics, all the results from this formalism agree with the
exact results, modulo nonperturbative terms. In the next
subsection, we use this formalism to derive the reflection coefficient and the
free energy of the theory and to describe the general perturbation of
the system.

\subsection{The reflection coefficient and the free energy}\label{sec3.2}
In the ${\cal S}$-matrix formulation of the ${\hat c}=1$ matrix
model, $\psi_-$ and $\psi_+$ may be regarded as incoming and
outgoing excitations, respectively. These in and out wave
functions are related by ${\cal S}$-matrix:
\begin{eqnarray}\label{transf1}
\psi_+(z_+,\bar{z}_+) = (S\psi_-) (z_+, \bar{z}_+) \,,
\end{eqnarray}
which is nothing but the Fourier transform given in
eq.~(\ref{Fourier2}).

The reflection coefficient $ {\cal R} (E, q)$ can be introduced in
the same way as the one in the $c=1$ matrix model described
earlier as
\begin{eqnarray}
(S\psi_-^{E, q})(z_+,\bar{z}_+) = {\cal R} (E, q) \psi_+^{E, q}(z_+,\bar{z}_+)~.
\end{eqnarray}
Straightforward computation leads to
\begin{equation}
{\cal R}(E, q)  =\frac{1}{2\pi}e^{\pi (E+i/2)} \Gamma(-iE
+\frac{q}{2}+\frac{1}{2} )\Gamma(-iE -\frac{q}{2}+\frac{1}{2} )\,.
\label{RefCoeff}
\end{equation}
Note that the exact expression of the reflection coefficient from
the deformed matrix model is given
by~\cite{Douglas:2003up,Demeterfi:1993cm}
\begin{equation}
{\cal R}(E, q)  =\frac{ \Gamma(-iE +\frac{q}{2}+\frac{1}{2} )}{\Gamma(iE
+\frac{q}{2}+\frac{1}{2} )}\,,
\label{RefCoeff1}
\end{equation}
which differs from the expression, (\ref{RefCoeff}), in only
nonperturbative corrections for $E$.

One can euclideanize the time coordinate in the target space and
consider the compactification of the ${\hat c}=1$ noncritical string
theory along that direction. In the context of the ${\hat c}=1$ matrix
quantum mechanics, the Euclidean version of the above reflection
coefficient can be obtained by replacing $E$ with $\mu + ip$.
Furthermore, since we are dealing with fermions which is
anti-periodic in compactified Euclidean time, the momentum modes
should be quantized as $p = \frac{n+1/2}{R}$ under
the compactification with the radius $R$. As alluded earlier, the
topological B-model on the deformed $\Z_2$ orbifolded conifold is
equivalent to the ${\hat c}=1$ 0A string theory compactified with
the radius $R=1$. Indeed, in the topological B-model we will obtain
the expression of the form~(\ref{RefCoeff}) after the replacement
$E\rightarrow \mu+ i(n+1/2)$.

One can obtain the `perturbative' free energy of the ${\hat c}=1$ theory from
the `perturbative' reflection coefficient~(\ref{RefCoeff}) in the
similar fashion as in the $c=1$ model outlined in the previous
section. Now $\phi_{0A}(E)=  Im\ln {\cal R}(E, q)$ is given by
\begin{equation}
\phi_{0A}(E) = {\rm Im} \Big[ \ln \Gamma
(-iE+\frac{q}{2}+\frac{1}{2})+\ln
\Gamma(-iE-\frac{q}{2}+\frac{1}{2})\Big]~,
\end{equation}
modulo irrelevant terms, or
\beq \frac{d^3\phi_{0A}(E)}{d E^3} = {\rm Im}
\Big[i\psi^{(2)}(-iE+\frac{q}{2}+\frac{1}{2})
+i\psi^{(2)}(-iE-\frac{q}{2}+\frac{1}{2})\Big]\,. \eeq
Therefore we have
\bea \frac{\partial^3}{\partial \mu^3} {\cal F}_{0A}(\mu, q) &=&
R~ {\rm I m }\bigg[ \int^{\infty}_{0}d t~ \Big\{e^{-i(\mu+iq/2)
t}+ e^{-i(\mu-iq/2) t}\Big\}\frac{t/2}{\sinh
(t/2)}\frac{t/(2R)}{\sinh (t/2R)}\bigg]
\nn  \\
&=& \frac{\partial^3 }{\partial \mu^3} \bigg[{\cal
F}_{c=1}\Big(\mu+\frac{i}{2} q\Big) +  {\cal
F}_{c=1}\Big(\mu-\frac{i}{2} q\Big)\bigg]~. \label{free}\eea
One may note that the free energy obtained from the deformed matrix
model is given by
\beq \frac{\partial^3}{\partial \mu^3} {\cal F}_{0A}(\mu, q) = 2R~
{\rm I m }\bigg[ \int^{\infty}_{0}d t~ e^{-i(\mu-iq/2) t}\,
\frac{t/2}{\sinh (t/2)}\frac{t/(2R)}{\sinh (t/2R)}\bigg]~, \eeq
which is the same as the above expression (\ref{free}) modulo
nonperturbative terms.

We can consider the general perturbation by momentum modes of
the tachyon field in the ${\hat c}=1$ 0A string theory. It was
suggested in \cite{Yin:2003iv} that it can be incorporated in the
matrix model by considering the new eigenfunctions
\begin{equation}\label{eigen}
\Psi_\pm^{E, q} = e^{\mp i\varphi_{\pm} (z_\pm\bar{z}_\pm ; E, q)}
\psi_\pm^{E, q}~,
\end{equation}
where the phases $\varphi_\pm$ have Laurent expansion
\beq \varphi_\pm (z_\pm\bar{z}_\pm ; E, q) = \frac{1}{2} \phi (E,
q)  + R\sum_{k\geq 1} t_{\pm k} (z_\pm\bar{z}_\pm)^{k/R}
-R\sum_{k\geq 1} \frac{1}{k}v_{\pm k} (z_\pm\bar{z}_\pm)^{-k/R}~.
\eeq

The ${\hat c}=1$ matrix model is the theory of two dimensional
free fermions with fixed angular momentum $q$, which corresponds
to the net D0-brane charge. Since the perturbation by tachyon
momentum modes preserves the background net D0-brane charge, it
should appear symmetrically in the Hamiltonian under $z_\pm
\leftrightarrow {\bar z}_\pm$. Therefore the perturbed Hamiltonian
in the ($z_+, \bar{z}_+$) representation may be given by
\begin{equation}
H_{tot}= H + z_+\frac{\partial\varphi_+}{\partial z_+}
+\bar{z}_+\frac{\partial\varphi_+}{\partial \bar{z}_+}~,
\end{equation}
whose eigenfunctions become (\ref{eigen}).
Later we will show that similar structure appears in the
deformation of complex moduli in the topological B-model on the
deformed $\Z_2$ orbifolded conifold.

\section{The Topological B model on the deformed orbifolded conifold}\label{sec4}
Now we are ready to study the topological B model on the
deformation of $\Z_2$ orbifolded conifold. The CY space we
consider is the hypersurface
\begin{equation}\label{hyper2}
uv- (xy-\mu_1)(xy-\mu_2)=0
\end{equation}
with the deformation parameters $\mu_1$ and $\mu_2$. In order to
have a non-singular geometry, we should have $\mu_1\neq \mu_2$.
Eventually we would like to show that this model is equivalent to
the $\hat{c}=1$ type 0A string theory with  the compactification
radius $R= \sqrt{\alpha'/2}$ in the background of net D0-brane charge
$q$. In this correspondence, the
deformation parameters are related to the cosmological constant
$\mu$ and the net D0-brane charge $q$ as
\begin{eqnarray}\label{mu_i}
\mu_1=g_s(\mu + \frac{i}{2}q)~, \qquad
\mu_2=g_s(\mu - \frac{i}{2}q)~.
\end{eqnarray}
The Riemann surface
\begin{eqnarray}\label{curve2}
H= (xy-\mu_1)(xy-\mu_2)=0
\end{eqnarray}
is given by the union of two sheets and each sheet corresponds to
genus zero surface with two boundaries.

In this section we describe the general set-up to solve the model.
In the next section we study the integrable structure of the model
and show the equivalence of various models.

\subsection{Orbifolded conifolds}\label{sec4.1}
In recent years, topological string theory on conifold has been
extensively studied~\cite{Aganagic:1999fe, dhot, otorb}. The
conifold is three dimensional singularity in $\C^4$ defined by
\bea uv -xy =0. \eea The conifold can be realized as a holomorphic
quotient of $\C^4$ by the $\C^*$ action given by
\cite{Witten:1993yc,Klebanov:1998hh}
\bea \label{act} (A_1, A_2,B_1, B_2)\mapsto (\lambda A_1, \lambda
A_2,\lambda^{-1} B_1, \lambda^{-1} B_2)\quad\mbox{ for }\lambda
\in \C^*. \eea
Thus it is a toric variety with a charge vector
$Q^{'}=(1,1,-1,-1)$ and the fan $\Delta=\sigma$ is given by  a
convex polyhedral cone in $\N^{'}_{\R}=\R^3$ generated by $v_1,
v_2, v_3, v_4 \in \N^{'}=\Z^3$  where
\bea v_1=(1,0,0), \quad v_2=(0,1,0),\quad v_3=(0,0,1),\quad
v_4=(1,1,-1). \eea

The isomorphism between the conifold ${\cal C}$ and the
holomorphic quotient is given by
\bea
 x=A_1B_1, \quad y=A_2B_2, \quad u=A_1B_2, \quad
v=A_2B_1. \eea
We take a further quotient of the conifold ${\cal C}$ by a
discrete group $\Z_k \times \Z_l$. Here $\Z_k$ acts on $A_i, B_j$
by
\bea \label{zk} (A_1, A_2, B_1, B_2) \mapsto (e^{-2\pi i/k} A_1,
A_2, e^{2\pi i/k}B_1, B_2), \eea and $\Z_l$ acts by \bea
\label{zl} (A_1, A_2, B_1, B_2) \mapsto (e^{-2\pi i/l} A_1, A_2,
B_1, e^{2\pi i/l}B_2). \eea
Thus they will act on the conifold ${\cal C}$ by
\bea \label{xy} (x,y,u,v) \mapsto (x,y,e^{-2\pi i/k}u, e^{2\pi
i/k}v) \eea and \bea \label{uv} (x,y,u,v) \mapsto (e^{-2\pi
i/l}x,e^{2\pi i/l}y, u, v). \eea

Its quotient is  called the hyper-quotient of the conifold or the
orbifolded conifold and denoted by ${\cal C}_{kl}$. To put the
actions (\ref{act}), (\ref{zk}) and (\ref{zl}) on an equal
footing, consider the over-lattice $\N$:
\bea \N = \N^{'} + \frac{1}{k}(v_3-v_1) + \frac{1}{l}(v_4 -v_1).
\eea
Now the lattice points $\sigma \cap \N$ of $\sigma$ in $\N$
 is generated by
$(k+1)(l+1)$ lattice points as a semigroup (These lattice points
will be referred as a toric diagram.). The charge matrix $Q$  will
be $(k+1)(l+1)$ by $(k+1)(l+1)-3$. The discrete group $\Z_k \times
\Z_l \cong \N / \N^{'}$ will act on the conifold $\C^4 // U(1)$
and its quotient will be the symplectic reduction $\C^{(k+1)(l+1)}
// U(1)^{(k+1)(l+1)-3}$ with the moment map associated with  the
charge matrix $Q$. The new toric diagram for ${\cal C}_{kl}$ will
also lie on the plane at a distance $1/\sqrt{3}$ from the origin
with a normal vector $(1,1,1)$.

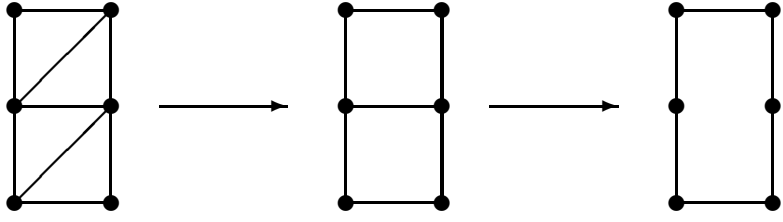
\begin{figure}
\setlength{\unitlength}{0.00083300in}%
\begingroup\makeatletter\ifx\SetFigFont\undefined%
\gdef\SetFigFont#1#2#3#4#5{%
  \reset@font\fontsize{#1}{#2pt}%
  \fontfamily{#3}\fontseries{#4}\fontshape{#5}%
  \selectfont}%
\fi\endgroup%
\begin{picture}(1000,1505)(2300,-2897)
\thicklines \put(3601,-2161){\circle*{100}}
\put(3601,-2761){\circle*{100}} \put(4201,-1561){\circle*{100}}
\put(4201,-2161){\circle*{100}} \put(4201,-2761){\circle*{100}}

\put(3601,-1561){\circle*{100}} \put(4201,-1561){\line(
0,-1){1200}} \put(3601,-1561){\line( 1, 0){600}}

 \put(3601,-1561){\line( 0,-1){1200}}
\put(3601,-2761){\line( 1, 0){600}}
\put(3601,-2761){\line(1,1){600}}
\put(3601,-2161){\line(1,1){600}}
 \put(3601,-2161){\line( 1,
0){600}} \put(4501,-2161){\vector(1,0){800}}
\end{picture}
\begin{picture}(1000,1505)(1300,-2897)
\thicklines \put(3601,-2161){\circle*{100}}
\put(3601,-2761){\circle*{100}} \put(4201,-1561){\circle*{100}}
\put(4201,-2161){\circle*{100}} \put(4201,-2761){\circle*{100}}

\put(3601,-1561){\circle*{100}} \put(4201,-1561){\line(
0,-1){1200}} \put(3601,-1561){\line( 1, 0){600}}

 \put(3601,-1561){\line( 0,-1){1200}}
\put(3601,-2761){\line( 1, 0){600}} \put(3601,-2161){\line( 1,
0){600}} \put(4501,-2161){\vector(1,0){800}}
\end{picture}
\begin{picture}(1000,1505)(300,-2897)
\thicklines \put(3601,-2161){\circle*{100}}
\put(3601,-2761){\circle*{100}} \put(4201,-1561){\circle*{100}}
\put(4201,-2161){\circle*{100}} \put(4201,-2761){\circle*{100}}

\put(3601,-1561){\circle*{100}} \put(4201,-1561){\line(
0,-1){1200}} \put(3601,-1561){\line( 1, 0){600}}

 \put(3601,-1561){\line( 0,-1){1200}}
\put(3601,-2761){\line( 1, 0){600}}
\end{picture}
\caption{Toric Diagrams
 {\bf left}: fully resolved geometry {\bf
middle}: partially resolved geometry {\bf right}:${\Z}_2$
orbifolded conifold}
\end{figure}

The action $(\ref{xy}), (\ref{uv})$ of $\Z_k \times \Z_l$ on the
conifold ${\cal C}$ can be lifted to an action on $\C^4$ whose
coordinates are $x,y, u, v$. The ring of invariants will be
$\C[x^l, y^l, xy, u^k, v^k, uv]$ and the orbifolded conifold
${\cal C}_{kl}$ will be defined by the ideal $(xy-uv)\C[x^l, y^l,
xy, u^k, v^k, uv]$. Thus after renaming variables, the defining
equation for the orbifolded conifold will be
\bea \label{con-eqn} {\cal C}_{kl}: xy =z^l, \quad uv =z^k. \eea

Hence the $\Z_2 \cong \Z_2 \times \Z_1 $ orbifolded conifold
${\cal C}_{k1}$ can be written as
\bea \label{z2orbifold} uv =(x y)^2.\eea
Its toric diagram is shown on the right of Figure 1. The general
complex deformation space of this singularity is given by the
Milnor ring \bea \frac{ \C \{ x, y, u, v\} }{(xy^2, x^2y, u,v )}
\cong \frac{ \C \{x, y\}}{(xy^2, x^2y)}.\eea As we stated before,
the ground ring of the $\hat{c}=1$ type 0A string theory is a
deformed $\Z_2$ orbifolded conifold
\begin{equation}\label{hyper22}
uv- \Big(xy-g_s(\mu + \frac{i}{2}q)\Big)\Big(xy-g_s(\mu -
\frac{i}{2}q)\Big)=0.
\end{equation}
After change of variables, one may rewrite this equation as
\bea u^2 + v^2 = (z-a)(z+a)~, \qquad x^2 +y^2 = z~, \qquad a>0\eea
and regard as a family over $z$-plane with generic fiber $\C^*
\times  \C^*$. By rescaling the variable $z$, we may assume that
$a$ is very closer to $0$. Then over the real segment $[0, a]$ the
family is like
\bea u^2 +v^2 \sim z-a, \quad \quad x^2 +y^2 \sim z\eea
which implies that \bea u^2 + v^2 -x^2 -y^2 \sim -a. \eea
Then $({\rm Im} \, u,  {\rm Im}\, v, {\rm Re}\, x, {\rm Re}\, y)$
describes an $\S^3$ cycle. Similarly over $[-a, 0]$, the family is
like
\bea u^2 +v^2 \sim -z-a, \quad \quad x^2 +y^2 \sim z\eea
which implies that
\bea u^2 + v^2 +x^2 + y^2 \sim -a. \eea
The quadruple $({\rm Im}\, u, {\rm Im}\, v, {\rm Im}\, x, {\rm
Im}\, y)$ describes another $\S^3$ cycle. So there two $S^3$
cycles.

As observed in~\cite{Hyun:2004yj}, the closed string theory on
this deformed conifold is dual to a open string theory on the
resolved conifold via a geometric transition through a partially
resolved conifold. In this situation, the partially resolved
conifold is obtained by introducing $\P^1$ which is shown on the
middle of the Figure 1.

Now the partially resolved conifold has two conifold singularities
and one can resolve these singularities by small resolution which
replaces each singularity by a $\P^1$ cycle. In the large $N$
duality, the $\S^3$ cycles on the deformed conifold are shrunken
and are replaced by the $\P^1$ cycles on the resolved conifold
whose toric diagram is shown on the left of Figure 1. So if one
considers the large $N$ duality of the open strings on the resolved
conifold, the
D-branes on $\P^1$'s disappear and the fluxes on $\S^3$ will be
generated in the closed string picture.

\subsection{Free energy}\label{sec4.2}
The free energy of the topological B-model on the CY geometry
given by~(\ref{hyper2}) can be obtained in the similar fashion as
in the case of the theory on the deformed conifold described in
section \ref{sec2.3}. The only difference is that now we have two
pairs of periods which correspond to two sheets in $H=0$. Again we
need to introduce a cut-off $\Lambda$ for the computation of
periods over noncompact $B_i$-cycle ($b_i$-cycle in $H=0$).  For
the curve
\begin{equation}
xy=\mu_i~,
\end{equation}
the periods can be computed as
\begin{eqnarray}
X^i &=& \oint_{A^i} \Omega = \frac{1}{2\pi i} \oint_{a^i} ydx = \mu_i\,, \nn \\
F_i &=& 2\pi i \int_{B^i} \Omega =
2\int_{\sqrt{\mu_i}}^{\sqrt{\Lambda}} y dx =\mu_i \ln
\Big(\frac{\Lambda}{\mu_i}\Big)~,
\end{eqnarray}
where the holomorphic three-form $\Omega$ is given by (\ref{3-form}).
Therefore the tree level free energy is given by the sum of two pieces as~(see also \cite{Danielsson:2004ti})
\begin{eqnarray}
{\cal F}_0 =\frac{1}{2}\sum_i X^i F_i= -\frac{1}{2} \sum_i
\mu_i^2\ln \mu_i +{\cal O} (\Lambda)~.
\end{eqnarray}

\begin{figure}
\begin{center}
    \begin{tabular}{cc}
      \includegraphics[scale=1]{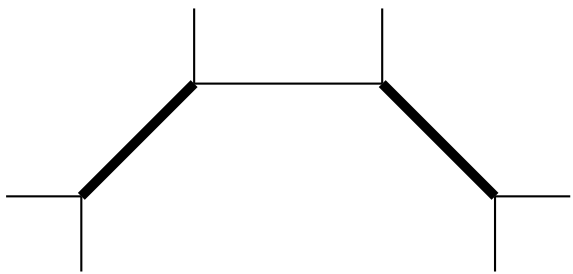}
    \end{tabular}
\caption{Diagram for fully resolved geometry where B-branes wrap
two $\P^1$ depicted as thick lines}\label{diagram}
\end{center}
\end{figure}

All order free energy of the topological B-model on the deformed
$\Z_2$ orbifolded conifold can be obtained by using the DV matrix
model. As alluded earlier, the deformed $\Z_2$ orbifolded conifold
has two $S^3$ cycles. RR-flux along two $S^3$ cycles can be
introduced without changing the topological
amplitudes~\cite{Vafa:2000wi}.  The closed topological B-model on
this deformed conifold is dual to the open topological B-model on
the resolved conifold via a geometric transition through a
partially resolved conifold.  The resolved geometry has three
$\P^1$, but the B-branes wrap along two disconnected $\P^1$ cycles
only (see Figure~\ref{diagram}). The open strings connecting
separated B-branes are massive and thus decoupled in the low
energy limit. Since the worldvolume theory of compact B-branes on
each $\P^1$ essentially reduces to the DV matrix model as given
by~(\ref{DV}), the whole theory may be described by the decoupled
$U(N_1)\times U(N_2)$ DV matrix model. The 't Hooft coupling $t_i $ of
each matrix model is identified with $i\mu_i/g_2$ in the
topological B-model on the deformed orbifolded conifold. Therefore we
have
\begin{equation}
{\cal F}_{top} (\mu_1, \mu_2) = {\cal F}_{DV}
\Big(t_1=i\frac{\mu_1}{g_s},~ t_2=i\frac{\mu_2}{g_s}\Big)= {\cal
F}_{DV} \Big(t_1=i\frac{\mu_1}{g_s}\Big)+ {\cal
  F}_{DV} \Big(t_2=i\frac{\mu_2}{g_s}\Big)~.
\end{equation}
Thus, using eqs. (\ref{DV1}) and (\ref{free}) with the identification as in (\ref{mu_i}), the all order
free energy of the topological B-model on the deformed orbifolded conifold
is identical with the free energy of the 0A string theory at the
compactification radius $R=1$:
\beq {\cal F}_{top} (\mu_1, \mu_2) = {\cal F}_{c=1}
\Big(\frac{\mu_1}{g_s}\Big)
  + {\cal F}_{c=1} \Big(\frac{\mu_2}{g_s}\Big)={\cal F}_{0A}
\Big(\mu=\frac{\mu_1+\mu_2}{2g_s},~
q=-i\frac{\mu_1-\mu_2}{g_s}\Big)~.
\eeq
This gives a strong indication that those two theories are indeed
equivalent.

\subsection{The structure of the curve}\label{sec4.3}

As argued in section~\ref{sec2}, the study of the topological
B-model on the CY space of the type (\ref{hyper1}) boils down to
the one of the complex deformations on the Riemann surface, $H=0$.
Furthermore, if the surface is given by the genus zero surface
with punctures, the complex deformations can appear only at the
`punctures'. In our case at hand, the geometry belongs to the CY
space of the type~(\ref{hyper1}) where the Riemann surface, $H=0$,
is given by the union of two curves
\begin{eqnarray}\label{curve3}
xy=\mu_1~, \ \ \ xy=\mu_2~.
\end{eqnarray}
Each curve describes a sphere with two punctures and hence the
Riemann surface can be regarded as the union of two spheres which
are `connected' at the punctures (see Figure 3). The region near
each puncture, or boundary can be associated with asymptotic
region described by $x\rightarrow \infty$ or $y\rightarrow \infty$
where the two curves `meet'.
%
%

\begin{figure}
\begin{center}
\begin{tabular}{cc}
      \includegraphics[scale=1]{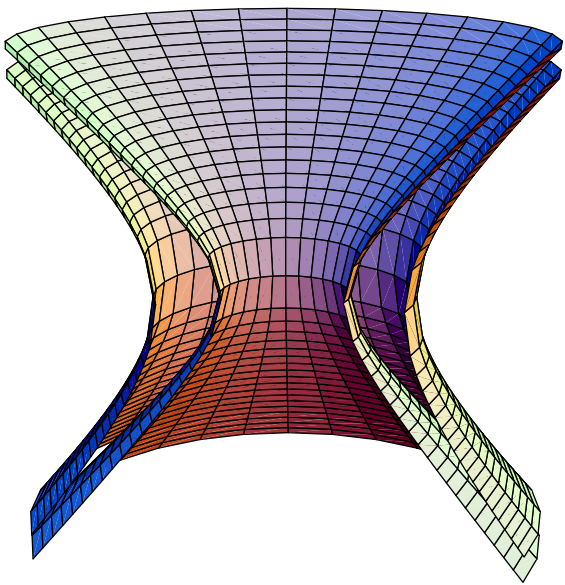}
\end{tabular}
\caption{Two-sheet Riemann surface relevant for type 0A string
theory}
\end{center}
\end{figure}

%
%
This tells us that we need to consider the complex structure
deformations on those two curves only. Furthermore the deformations
can appear only at the boundaries where two curves are
`connected'. Therefore the complex deformations at the boundaries
influence both curves at the same time. Those deformations near the
boundaries, $x\rightarrow \infty$ and $y\rightarrow \infty$, are
generically described, respectively, by
\begin{eqnarray}
\delta y &=& -g_s\sum_{n=1}^{\infty}n t_n x^{n-1} + g_s
\sum_{n=1}^{\infty}\frac{\partial}{\partial t_n} x^{-n-1}\,, \nn
\\ &&
\label{deform1} \\
\delta x &=&
-g_s\sum_{n=1}^{\infty}n\tilde{t}_ny^{n-1} + g_s
\sum_{n=1}^{\infty}\frac{\partial}{\partial \tilde{t}_n}
y^{-n-1}\,. \nn
\end{eqnarray}

In order to describe these complex deformations (\ref{deform1}) on
the surface (\ref{curve2}), it is convenient to introduce
independent coordinates $x_i, y_i, i=1,2$ for each curve in
(\ref{curve3}) and denote each curve as
\begin{equation}
 H_i(x_i, y_i) = x_i y_i-\mu_i=0\,.
\end{equation}
In this description, we study the complex deformations on the
curves $H_i=0$ which become those deformations
(\ref{deform1}) on the curve $H=0$ after the identifications of
the coordinates, $x_1=x_2=x$, $y_1=y_2=y$. Then the complex
deformations described above are those which deform the curves
$H_i =0$, while $H_1-H_2$ fixed.

Alternatively, one may begin with the higher dimensional geometry with
\begin{equation}
uv-H_1(x_1, y_1)H_2(x_2, y_2) =0~,
\end{equation}
where $H_i = x_i y_i-\mu_i$, $i=1,2$.
 This is a local Calabi-Yau fivefold, which can be regarded as
 Calabi-Yau threefold for fixed $x_1$ and $y_1$ (or $x_2$ and
 $y_2$). The geometry we consider corresponds to the subspace with the
 identification $x_1=x_2$ and $y_1=y_2$.
Then one can study the complex deformations on the curves $H_i=0$
which again give the complex deformations on the curve $H=0$
after the identifications. 
Only the symmetric combinations of $x_1$ and $x_2$ ($y_1$
and $y_2$) among the complex deformations of the curve 1 and 2
survive under the identifications. This is due to $\Z_2$ symmetry
under $x_1\leftrightarrow x_2$, $y_1\leftrightarrow y_2$ and
$\mu_1\leftrightarrow \mu_2$, which is inherited from the original
$\Z_2$ symmetry  under $\mu_1\leftrightarrow \mu_2$ of the CY
space~(\ref{hyper2}). In this way, the
relation between the ${\hat c}=1$ 0A string theory and the topological
B-model on the deformed orbifolded conifold  becomes mostly clear.

Indeed, one may regard $H_1$ and $H_2$ as Hamiltonians for two
sheets, and they correspond to the  $z_+$ and
$\bar{z}_+$ part of the Hamiltonian in the free fermionic description of the type 0A string theory. We can combine these Hamiltonians as
\begin{eqnarray}\label{ham2}
\tilde{H}&=& \frac{1}{2}(H_1+H_2) \nn \\
&=&   x_1 y_1+ x_2 y_2-\mu_1-\mu_2\,,
\end{eqnarray}
and
\begin{eqnarray}\label{ang2}
J&=& (H_1-H_2) \nn\\
&=&   x_1 y_1- x_2 y_2-\mu_1+\mu_2\,.
\end{eqnarray}
The similarity between eq.s~(\ref{ham1}, \ref{ang1}) and
eq.s~(\ref{ham2}, \ref{ang2}) is quite striking! Indeed, we obtain
the complex deformations of the surface (\ref{curve2}) by
restricting the deformations of $H_i=0$ to the deformations of
$\tilde{H}=0$ while $J=-\mu_1+\mu_2$ fixed.

\subsection{Free fermion description}\label{sect4.4}

As in the deformed conifold case, we can parametrize the complex
deformations as Laurent expansions of $y_i(x)$ in the $x_i$-patch
and $x_i(y)$ in the $y_i$-patch by
\begin{eqnarray}
y_i=\partial_{x_i}\phi(x_1,x_2)\,, \qquad
x_i=-\partial_{y_i}\tilde{\phi}(y_1,y_2)\,.
\end{eqnarray}
The classical part of $\phi$ is given by
\begin{eqnarray}
\partial_{x_i}\phi_{cl}(x_1,x_2)=\frac{\mu_i}{x_i}\,,
\qquad  \partial_{y_i}\tilde{\phi}_{cl}(y_1,y_2) =
-\frac{\mu_i}{y_i}\,.
\end{eqnarray}
As noted above, the relevant complex deformations are those which
survive after the identification, or those which satisfy
$J=-\mu_1+\mu_2$. Therefore the appropriate complex deformations are
\begin{eqnarray}\label{deform}
\phi(x_1,x_2)&=& \mu_1\ln x_1 +\mu_2\ln x_2 - g_s
\sum_{n=1}^{\infty}t_n(x_1x_2)^n - g_s
\sum_{n=1}^{\infty}\frac{1}{n}\frac{\partial}{\partial t_n}
(x_1x_2)^{-n}\,, \nonumber \\ && \\
\tilde{\phi}(y_1,y_2)&=& -\mu_1\ln y_1 - \mu_2\ln y_2 - g_s
\sum_{n=1}^{\infty}\tilde{t}_n(y_1y_2)^n  - g_s
\sum_{n=1}^{\infty}\frac{1}{n}\frac{\partial}{\partial \tilde{t}_n}
(y_1y_2)^{-n}\,, \nonumber
\end{eqnarray}
where, in the classical limit, the couplings are simply given by periods
\bea g_s t_n &=& -\frac{1}{n} \oint_{x_1\rightarrow \infty}
\oint_{x_2\rightarrow \infty} y_1y_2 (x_1 x_2)^{-n}\,, \nn \\
g_s \frac{\partial {\cal F}_0}{\partial t_n} &=& \oint_{x_1\rightarrow
\infty} \oint_{x_2\rightarrow \infty} y_1y_2 (x_1 x_2)^{n}\,. \eea
Each term in $t_n$ corresponds to the deformation $y\rightarrow y
+ nx^{2n-1}$ in the deformed $\Z_2$ orbifolded conifold.

Let us consider the deformation of the geometry by inserting
non-compact B-branes wrapping the fiber i.e. $u, v$ directions. As
we described earlier in the previous subsection, the deformation
of the geometry can appear only at the asymptotic region and
affect both curves in the same amount. Henceforth, we may consider
the B-branes at the boundary affecting both  curves simultaneously. In the higher dimensional picture, this corresponds to
putting B-branes at the boundaries of both curves, $H_i=0$, at the
same time. From the brane worldvolume theory, one may regard
($x_i, y_i$) as conjugate pairs, and therefore they play dual roles
as a coordinate in one patch and momentum in the other, and vice
versa.

In the closed string
picture, these B-branes can be incorporated by the brane
creation/annihilation operators with two complex
variables:
\beq \psi(x_1,x_2) =~ :\exp
\Big[-\frac{i}{g_s}\,\phi(x_1,x_2)\Big]:~, \quad
\tilde{\psi}(y_1,y_2) =~ :\exp
\Big[-\frac{i}{g_s}\,\tilde{\phi}(y_1,y_2)\Big]:~, \eeq
which correspond to the deformations of the geometry with (\ref{deform}).
Since $x_i$ and $y_i$ are conjugate each other, the transformation law
between fermions in $(x_1,x_2)$ and $(y_1,y_2)$
patches is given by the Fourier transform:
\begin{equation}\label{trans}
\tilde{\psi}(y_1,y_2) = \frac{1}{2\pi} \int dx_1dx_2~
e^{\frac{i}{g_s}(y_1x_1 +y_2x_2)}\, \psi(x_1,x_2)\,. \label{TSFTR}
\end{equation}
The fermions $\psi(x_1,x_2)$ in $x_i$ patch and
$\tilde{\psi}(y_1,y_2)$ in $y_i$ patch may correspond to {\it in} and {\it out}
states, respectively, in the ${\cal S}$-matrix formulation of ${\hat c}=1$ theory.

It is convenient to perform the following change of
variables\footnote{One may choose $z'= x_2$ and $w'=y_2$, instead.
This gives the same results due to the $\Z_2$ symmetry of the
model under $\mu_1 \leftrightarrow\mu_2$ (or $q\leftrightarrow -q$). }
\begin{eqnarray}
z=x_1x_2\,, \ \ \ z' = x_1\,, \ \ \  w=y_1y_2\,, \ \ \ w'= y_1\,,
\label{CVar}
\end{eqnarray}
and to introduce new fermionic functions $\chi$ and $\chi_{qu}$,
which will be used repeatedly, as follows:
\begin{eqnarray}
\psi(x_1, x_2) &=& {z'}^{i(\mu_2-\mu_1)/g_s}\chi(z)\, =\,
{z'}^{i(\mu_2-\mu_1)/g_s}z^{-i\mu_2/g_s}\chi_{qu}(z)\,, \nonumber
\\ && \label{Chi}\\ \tilde{\psi}(y_1, y_2) &=&
{w'}^{i(\mu_1-\mu_2)/g_s}\tilde{\chi}(w)\,=\,
 {w'}^{i(\mu_1-\mu_2)/g_s}
w^{i\mu_2/g_s}\tilde{\chi}_{qu}(w)\,. \nn
\end{eqnarray}
The fermion field $\chi_{qu}$ represents the quantum part of the
brane creation operator. As it depends only on one variable
$z=x_1x_2$~(or $w=y_1y_2$), the treatment of this function is
analogous to the one in the case of the deformed conifold, where the
Riemann surface is given by single-sheet with two punctures~(see
section~\ref{sec2.3}). As explained earlier, the deformation of the
Riemann surface should affect the two sheets at the same time in
the same way, thus it is natural that the brane/fermion creation
operator $\chi_{qu}$ depends only on the symmetric combination of
$x_1$ and $x_2$, and thus $z=x_1x_2$.

Therefore the brane/fermion creation operator $\chi_{qu}$ can be mode-expanded as
\beq \chi_{qu}(z) = \sum_{n\in \Z}\,
\chi_{n+\frac{1}{2}}z^{-n-1}\,.\eeq
Also for later convenience, we introduce $\phi_{\chi}$ and
$\phi_{qu}$ which come from the bosonization of $\chi$ and
$\chi_{qu}$, respectively, as follows:
\bea && ~ \chi(z) =~ :\exp\Big[-\frac{i}{g_s}\,
\phi_{\chi}(z)\Big] :~, ~~~\qquad \phi_{\chi}(z) =
\mu_2\ln z + \phi_{qu} (z)\,, \\
&&\chi_{qu}(z) =~ :\exp\Big[ -\frac{i}{g_s}\,\phi_{qu}(z)\Big] :~,
 \qquad \phi_{qu}(z) = -g_s\sum_{n=1}^{\infty}t_nz^n -g_s
\sum_{n=1}^{\infty}\frac{1}{n}\frac{\partial}{\partial t_n}
z^{-n}\,. \nn \eea
The transformation law between fermions $\chi$
in $x_i$ patch and $\tilde{\chi}$ in $y_i$ patch can be read from~(\ref{TSFTR}):
\beq \label{trans5} \tilde{\chi}(w) =\frac{1}{2\pi}\int dz \int
\frac{ds}{s}~ s^{\frac{i}{g_s}(\mu_2-\mu_1)}\,
e^{\frac{i}{g_s}(s+zw/s)}~ \chi(z)\,. \eeq
%

\section{Integrable structure of topological B-model}\label{sec5}
In this section we describe the integrable structure of the
topological B-model on the deformation of $\Z_2$ orbifolded
conifold using the fermionic description introduced in the
previous section. First of all, we describe the ${\cal W}$ algebra
and the associated Ward identity which is identical with the one
in the ${\hat c}=1$ 0A string theory compactified at the radius $R=1$.
Then we obtain the state $|V\rangle$ which corresponds to the
quantum free energy and reproduce the reflection coefficient of
the corresponding ${\hat c}=1$ string theory. Finally we obtain
the Kontsevich-like matrix model which is related to the
topological B-model on the deformed orbifolded conifold and the ${\hat
c}=1$ string theory.

\subsection{Ward identities}\label{sec5.1}

We can solve this topological B-model using the symmetry
underlying the geometry, in the similar fashion as in the case of the
deformed conifold. To this end, let us consider the operator
\[\oint_P\oint_P (x_1 x_2)^n
\partial_{x_1} \partial_{x_2} \phi(x_1, x_2)~. \] It corresponds
to the ${\cal W}$-symmetry generator which gives the
transformation
\[
(x, y) \rightarrow (x, y + n x^{2n-1})~.
\]
The associated Ward identity can be read as
\begin{eqnarray}
&&\oint_{x_1\rightarrow \infty} \oint_{x_2\rightarrow \infty}
\psi^*(x_1, x_2) (x_1x_2)^n
\psi(x_1, x_2) |V\rangle = \nonumber \\
&& ~~~~~ = - \oint_{y_1\rightarrow \infty} \oint_{y_2\rightarrow
\infty} \tilde{\psi}^*(y_1, y_2)
\Big((-ig_s)^2\partial_{y_1}\partial_{y_2}\Big)^n
\tilde{\psi}(y_1, y_2) |V\rangle\,.
\end{eqnarray}

It is convenient to perform the change of variables as in
(\ref{CVar}), under which the differential operator becomes
\beq g_s^2\partial_{y_1}\partial_{y_2} =g^2_s\Big[
w^2\frac{\partial^2}{\partial w^2} + \frac{\partial}{\partial w} +
w'\frac{\partial^2}{\partial w\partial w'}\Big]\,. \eeq
After the trivial contour integration over $z'$ and
$w'$, we obtain
\begin{eqnarray}\label{Ward4}
&& \oint_{z\rightarrow \infty} \chi^*(z) z^n \chi(z) |V\rangle
=\nonumber \\&& ~~~~~= - \oint_{w\rightarrow \infty}
\tilde{\chi}^*(w) (-g_s^2)^{n}\Big[w\partial_{w}^2
+\Big(1+i\frac{\mu_1-\mu_2}{g_s}\Big)\partial_{w}\Big]^n
\tilde{\chi}(w) |V\rangle\,.
\end{eqnarray}
This shows that, in the computation of the Ward identity, two
coordinates $x_i$, are effectively reduced to one coordinate $z$,
which corresponds to $x^2$ after the identification.

One may note that
\begin{eqnarray}
\widetilde{\cal W}_n \equiv \oint_{w\rightarrow \infty}
\tilde{\chi}^*(w) (-g_s^2)^{n}\Big[w\partial_{w}^2
+\Big(1+i\frac{\mu_1-\mu_2}{g_s}\Big)\partial_{w}\Big]^n
\tilde{\chi}(w) \,
\end{eqnarray}
commute among themselves and form Cartan subalgebra of ${\cal
W}_{1+\infty}$
algebra~\cite{Pope:1990rn,Sezgin:1990ee,Fukuma:1990yk}. They are
responsible for the integrability of the topological B-model on
the deformed orbifolded conifold as for the integrability of the
${\hat c}=1$ 0A string theory.

Since
\begin{eqnarray}\label{deriv}
&&(w\partial_{w}^2 +A\partial_{w})^n =\sum_{k=0}^{n} {n \choose k}
[A+n-1]_k w^{n-k} \partial_w^{2n-k}\,,
\end{eqnarray}
where
\begin{equation}
[x]_k  \equiv x(x-1) \cdots (x-k+1)\,, \qquad [x]_{0}=1\,,
\end{equation}
the Ward identity~(\ref{Ward4}) can be expressed as
\begin{eqnarray}\label{ward3}
&& \oint_{z\rightarrow \infty}d z~ \chi^*(z) z^n
\chi(z) |V\rangle ~~
 =
~~~~~~~~~\\ && ~~~~~ =~ -(-g_s^2)^{n}\sum_{k=0}^{n} {n \choose k}
[A+n-1]_k \oint_{w\rightarrow \infty}d w~ w^{n-k}~
\tilde{\chi}^*(w)
\partial_w^{2n-k}\tilde{\chi}(w)\, |V\rangle\,, \nonumber
\end{eqnarray}
where $A=1+i(\mu_1-\mu_2)/g_2=1-q$.

In the ${\cal S}$-matrix formulation\cite{Aganagic:2003qj} with coherent state basis,
the perturbed partition function is given by
\beq  Z(\tilde{t}_{n},t_n) \equiv \langle {\tilde t}|V\rangle= -\langle \tilde {t}| S |t \rangle\,,
\qquad |t\rangle \equiv
e^{\frac{i}{g_s}\sum_{n=1}^{\infty}t_n\tilde{\alpha}_{-n}}|0\rangle\,.
\eeq
This leads us to the alternative form of the Ward identity:
\bea \label{ward4} \frac{1}{i} \frac{\partial }{\partial
t_n}Z(\tilde{t}_n,t_n) &=& -(-g_s^2)^n \sum_{k=0}^{n}{n\choose
k}[n-q]_k \frac{1}{2n+1-k}~ ~ \times  \\&&~~~~ \times~  \oint
\frac{d w}{2\pi i}~ w^{n-k} :
e^{\frac{i}{g_s}\tilde{\phi}_{\chi}(w)}\partial_w^{2n+1-k}
e^{-\frac{i}{g_s}\tilde{\phi}_{\chi}(w)}:~ Z(\tilde{t}_n,t_n)\,.
\nn \eea
This is exactly the same as the Ward identity which appears in the
generating functional of the tachyon momentum mode perturbation in the
type 0A theory at the compactification radius $R=1$, given
in~\cite{Ita:2004yn}.

\subsection{The quantum free energy and the reflection coefficient}\label{sec5.2}

As explained earlier, the quantum free energy of the topological
B-model can be represented by the state $|V\rangle$ in the Hilbert
space ${\cal H}^{\otimes k}$. If there is no deformation at the
boundaries, the state $|V\rangle$ is given by the fermionic vacuum
state. After the deformations by a set of  $t^i_m$, the state
$|V\rangle$ may be given by the Bogoliubov transformation of the
fermionic vacuum by the quantum part of fermionic operators
\beq |V\rangle = \exp\Big[
\sum_{m,n\geq 0}a_{mn}\chi_{-m-\frac{1}{2}}\tilde{\chi}^*_{-n-\frac{1}{2}}
+\sum_{m,n\geq 0 }
\tilde{a}_{mn}\tilde{\chi}_{-m-\frac{1}{2}}\chi^*_{-n-\frac{1}{2}}\Big]
|0\rangle\,. \eeq
As it turns out, the coefficients of the Bogoliubov transformation
correspond to the reflection coefficients of the ${\hat c}=1$ theory.

In order to determine $|V\rangle$, let us consider the two point
function $\langle 0|\tilde{\psi}(y_1,y_2)$
$\psi^*(x_1,x_2)|V\rangle $. As usual, the transformation law
between fermions of ($x_1, x_2$) and ($y_1, y_2$) patches is given
by the Fourier transform,~(\ref{trans}) and thus the two-point
function can be written as
\begin{equation}\label{2-point}
\langle 0|\tilde{\psi}(y_1,y_2) \psi^*(x_1,x_2)|V\rangle =
\frac{1}{2\pi}\int du_1du_2~ e^{\frac{i}{g_s}(y_1u_1 +
y_2u_2)}\, \langle 0|
\psi(u_1,u_2)\psi^*(x_1,x_2)|V\rangle\,.
\end{equation}

This two point function can be computed in two different ways. By
using $\chi$ introduced in (\ref{Chi}), the left hand side of
 (\ref{2-point}) can be expressed as
\begin{eqnarray}\label{left}
\langle 0|\tilde{\psi}(y_1,y_2) \psi^*(x_1,x_2)|V\rangle &=&
(w'z')^{\frac{i}{g_s}(\mu_1-\mu_2)} (w
z)^{\frac{i}{g_s}\mu_2}\langle 0|\tilde{\chi}_{qu}(w)
\chi^*_{qu}(z)|V\rangle \\ \nonumber
&=&(x_1y_1)^{\frac{i}{g_2}\mu_1}(x_2y_2)^{\frac{i}{g_2}\mu_2}
\sum_{m,n\ge 0}a_{mn}(x_1x_2)^{-m-1}(y_1y_2)^{-n-1}~.
\end{eqnarray}
On the other hand, by using the standard operator product
expansion of $\chi$, which is the function of one variable $z$,
the two point function in the right hand side of (\ref{2-point})
can be computed as
\begin{eqnarray}
\langle 0| \psi(u_1,u_2)\psi^*(x_1,x_2)|V\rangle &=&
\Big(\frac{z'}{u'}\Big)^{\frac{i}{g_s}(\mu_1-\mu_2)}
\Big(\frac{z}{u}\Big)^{\frac{i}{g_s}\mu_2}\langle 0|\chi_{qu}(u)
\chi^*_{qu}(z)|V\rangle \nonumber
\\
&=&\Big(\frac{z'}{u'}\Big)^{\frac{i}{g_s}(\mu_1-\mu_2)}
\Big(\frac{z}{u}\Big)^{\frac{i}{g_s}\mu_2} \frac{1}{u-z}\,,
\end{eqnarray}
where $u\equiv u_1u_2$ and $u'\equiv u_1$.  After performing an expansion over $\frac{u}{z}$, the right hand side
of  (\ref{2-point}) becomes
\begin{eqnarray}\label{right}
&& -(x_1y_1)^{\frac{i}{g_2}\mu_1}(x_2y_2)^{\frac{i}{g_2}\mu_2}
\sum_{n=0}^{\infty}(x_1x_2)^{-n-1}(y_1y_2)^{-n-1}  ~~ \times  \nn \\
&&  ~~~~~~~~~~~~~~~\times~~ \frac{1}{2\pi}\int d\eta'~
e^{\frac{i}{g_s}\eta'}{\eta'}^{-\frac{i}{g_s}\mu_1} {\eta'}^n \int
d\eta~ e^{\frac{i}{g_s}\eta}\,\eta^{-\frac{i}{g_s}\mu_2} \eta^n\,.
\end{eqnarray}
By comparing both sides of (\ref{2-point})  from (\ref{left}) and
(\ref{right}), the coefficients of Bogoliubov transformation can
be determined as $a_{mn}= -R_n\delta_{mn}$ and $\tilde{a}_{mn}=-
R_n^*\delta_{mn}$ where
\beq R_{n} = \frac{1}{2\pi}\int d\eta'~
e^{\frac{i}{g_s}\eta'}{\eta'}^{-\frac{i}{g_s}\mu_1} {\eta'}^n \int
d\eta~ e^{\frac{i}{g_s}\eta}\,\eta^{-\frac{i}{g_s}\mu_2} \eta^n\,.
\eeq
By integrating along the negative real axis, just as we did in the ${\hat c}=1$
case, we obtain the expression for the coefficient $R_n$ as
\beq R_{n} = \frac{1}{2\pi}
e^{\pi\mu}e^{i\pi(n+1)}\Gamma\Big(-i\mu +n + 1 +\frac{q}{2}\Big)
\Gamma\Big(-i\mu +n + 1 - \frac{q}{2}\Big)\,. \eeq
It is identical with the Euclidean version of `the perturbative'
reflection coefficient of the type 0A theory at the radius $R=1$, given in
(\ref{RefCoeff}).

\subsection{Kontsevich-like matrix model}\label{sec5.3}
In this section we derive the Kontsevich-like matrix model
corresponding to the ${\hat c}=1$ 0A string theory. Let us consider
the deformation of the geometry through only at one boundary, say
$x\rightarrow\infty$. In this situation, only $t_n$, the
deformation parameters in one patch, are turned on, while those in
the other patch, ${\tilde t}_n$ remain zero. All the amplitudes
remain trivial as far as $\tilde{t}_n=0$, which can be easily
understood from the fact that the corresponding amplitudes in the
${\hat c}=1$ 0A string theory vanish due to momentum conservation.

Now we put $N$ non-compact B-branes at positions $y_1=y_{1,l}$ and
$y_2=y_{2,l}$, $l= 1,\cdots, N$ near the boundary $y_i\rightarrow
\infty$. As explained earlier, this deforms the background
geometry with deformation parameters given by
\beq\label{Miwa}
 \tilde{t}_n =  \frac{i}{n} {\rm Tr}\, A^{-n}\,,
 \qquad A={\rm diag}(w_1,w_2,\cdots,w_N),
 \eeq
where $w_l=y_{1,l}y_{2,l}$ as introduced in~(\ref{CVar}).

The perturbed partition function in terms of new variables $w_l$ and $w'_l$ as
in (\ref{CVar}) can be written as
\begin{eqnarray}
Z(\tilde{t}_n,t_n) &=& \frac{1}{\Delta (w)} \langle N| \prod_{l=1}^N
\tilde{\chi}_{qu} (w_l) | V \rangle \nn \\
&=& \frac{(\det A)^{-\frac{i}{g_s}\mu_2}}{\Delta (w)}  \langle N| \prod_{l=1}^N
\tilde{\chi} (w_l) | V \rangle~,
\end{eqnarray}
where $|V\rangle$ is the state corresponding to the deformed
geometry with $t_n$ only and $|N\rangle$ denotes the $N$ fermion
state. The normalization can be determined from the condition $Z(\tilde{t}_n,
t_n)|_{t_n=0}=1$.

In order to determine this perturbed partition function, first of
all, we perform the transformation from $\tilde{\chi}$ to $\chi$
 using (\ref{trans5}). Then by using (\ref{prod2}), we obtain
\begin{equation}
\langle N| \chi(z_1)\chi(z_2)\cdots
\chi(z_N)|V\rangle = \Delta (z)~
e^{\frac{i}{~g_s}\sum_{l=1}^N
\big(-\mu_2\ln z_l+g_s\sum_{n>0}t_nz_l^n\big)}\,.
\end{equation}
By collecting pieces together, the partition function can be expressed as
\bea \label{part6}Z(\tilde{t}_n,t_n)  &=& \frac{(\det
A)^{-\frac{i}{g_s}\mu_2}}{\Delta (w)}  \int \prod_{l=1}^N
\frac{dz_l}{2\pi}~  \Delta (z)\, e^{\frac{i}{g_s}\sum_{l}\Big\{-
\mu_2\ln z_l + g_s\sum_{n>0} t_nz^n_l\Big\}} ~~\times  \nn
\\ &&~~~~~ \times~~  \int \prod_{l=1}^N ds_l~
s_l^{q-1} \, e^{\frac{i}{g_s}\sum_{l>0}\big\{s_l +
w_lz_l/s_l\big\}}\,. \eea

In this form, it is straightforward to show that the perturbed
partition function satisfies the Ward identity. Using the brane
position variables~(\ref{Miwa}) which has been known as
the Miwa-Kontsevich transform~\cite{Dijkgraaf:1991qh,Mukhi:2003sz} and
performing the residue integrals analogous to the $c=1$ matrix model
case~\cite{Imbimbo:1995yv}, the Ward identity~(\ref{ward4})
becomes
\bea  \!\!\! \!\!\!  \frac{1}{i}\frac{\partial Z(\tilde{t}_n,t_n)
}{\partial t_n}&=& (-g_s^2)^n \sum_{k=0}^{n}{n\choose k}[n-q]_k
\bigg\{ \sum_{l=1}^{N}\frac{w_l^{-i\mu_2/g_s}}{\prod_{m\neq l}
(w_m-w_l)} ~~\times  \nn
\\  \!\!\!  \!\!\! && ~~~~~~~~~  \times~      w_l^{n-k} \Big(\frac{\partial}{\partial
w_l}\Big)^{2n-k}  w_l^{i\mu_2/g_s} \prod_{m\neq l}(w_m-w_l)
\bigg\} ~ Z(\tilde{t}_n,t_n)  \nn \\ \!\!\! \!\!\!   &=&
(-g_s^2)^n (\det A)^{-i\frac{\mu_2}{g_s}}~ \sum_{l=1}^{N}~
\frac{1}{\Delta (w)}
 ~~ \times  \nn \\
&& ~~~~~ \times ~
 \Big[w_l\frac{\partial^2}{\partial w_l^2}+
(1-q)\frac{\partial}{\partial w_l}\Big]^n \Delta (w)~ (\det
A)^{i\frac{\mu_2}{g_s}}~ Z (\tilde{t}_n,t_n)\,. \label{WI3}  \eea

Note that the perturbed partition
function in (\ref{part6}) contains the function $F_q(x)$ defined as
\beq F_q(x) =  \int d s~ s^{q-1} \, e^{\frac{i}{g_s}(s + x/s)}~,
\eeq
which satisfies the Bessel equation:
\beq g^2_s[w\partial^2_w + (1-q)\partial_w] F_q(w z) = - z F_q(w z)\,.
\eeq
This function can be written in terms of (modified) Bessel (or
Hankel) functions as $F_{\nu}(x) =
x^{\nu/2}Z_{\nu}(2\sqrt{x}/g_s)$ up to constant. If the
integration region is taken over the negative real axis, $Z_{\nu}$
is given by the Hankel function $H_{\nu}^{(2)}$. This shows clearly
that the perturbed partition function~(\ref{part6}) satisfies the
Ward identity of the form~(\ref{WI3})\,.

Through the change of variables $s_l\rightarrow w_ls_l$,  the
perturbed partition function~(\ref{part6}) can be rewritten as
\begin{eqnarray}
Z(\tilde{t}_n,t_n) &=& \frac{(\det
  A)^{-\frac{i}{g_s}\mu_1}}{\Delta (w)}\int \prod_{l=1}^N \frac{ds_l}{s_l}~
e^{\frac{i}{g_s}\sum_{l}\Big\{s_lw_l  -
(\mu_1-\mu_2)\ln s_l\Big\}} ~~ \times \nonumber \\
&& ~~~~~ \times~~  \int \prod_{l=1}^N dz_l~  \Delta (z)\,
e^{\frac{i}{g_s}\sum_{l}\Big\{(z_l/s_l) - \mu_2\ln z_l +
g_s\sum_{n>0} t_nz^n_l\Big\}}\,.
\end{eqnarray}
After using the Harish-Chandra-Itzykson-Zuber-Mehta
integral\cite{Harish-Chandra,Itzykson:1979fi,Mehta:1981xt}
\begin{equation}
\int dU~\, e^{i\,{\rm Tr}\, (UXU^{\dagger}Y)} = {\rm const.}\, \cdot\, \frac{
\det e^{ix_iy_j}}{\Delta (x) \Delta(y)}\,,
\end{equation}
and
\begin{equation}
\Delta\Big(\frac{1}{s}\Big) = \frac{\Delta(s)}{~
\prod_{l}s_l^{N-1}}\,,
\end{equation}
one can see that the perturbed partition function becomes the one
of Kontsevich-like matrix model:
\bea Z(\tilde{t}_n,t_n) & =& (\det A)^{-\frac{i}{g_s}\mu_1}\int
{\cal D}S~ e^{\frac{i}{g_s}\big\{\Tr(AS) - (\mu_1-\mu_2 - ig_s
N)\Tr\ln S \big\}} ~~\times \nonumber \\
&& ~~~~~~~~ \times~~ \int {\cal D}Z~
e^{\frac{i}{g_s}\big\{\Tr(S^{-1}Z) -\mu_2\Tr\ln Z +g_s\sum_{n > 0}
t_n \Tr Z^n\big\}}~.  \eea
Note that by inserting the B-branes at the boundary the
deformation parameter $\mu_i$ is shifted to
$\mu_i-ig_sN$~\cite{Aganagic:2003qj}. After this shift, it is the
same Kontsevich-like matrix model as the one dual to the ${\hat
c}=1$ type $0A$ theory.

\section{Conclusion}
In this paper we studied the topological B-model on the
deformation of $\Z_2$ orbifolded conifold and show that it is
equivalent to the ${\hat
  c}=1$ theory compactified at the radius  $R=\sqrt{\alpha'/2}$ with
nonzero background D0-brane charge.
 Via the DV
matrix model, we obtained the B-model free energy and showed it is identical with the ${\hat c}=1$ free energy.
Most notably, we showed that the topological B-model on that geometry
admits free fermion description, exactly the same way as the 
${\hat c}=1$ theory. Furthermore we derived the Ward identities of the
model. This led us to obtain the perturbed free energy of the
model and the corresponding Kontsevich-like matrix model. All these
confirm the equivalence between those two models.

While in the ${\hat c}=1$
theory side, $q=0$ case is well-defined, the corresponding case
in the topological B-model is not well-defined due to non-isolated
singularities. Nevertheless one may notice that  all the quantities we obtained in the
topological B-model are
smooth in the limit $\mu_1\rightarrow \mu_2$ ($q\rightarrow 0$). This
suggests that the equivalence of those two theories in the $q=0$ case
should be understood as a limiting process. Namely, in the topological
B-model side, we first regularize the singularity of the geometry by the
infinitesimal deformation with $\delta =\mu_1-\mu_2$ and then take the
smooth limit $\delta\rightarrow 0$.

In the compactified string theory, we also need to consider the
winding modes. Indeed the perturbation by winding modes only is
also integrable~\cite{Kazakov:2000pm,Park:2004yc}. But it is not
clear in the context of free fermion description how to
incorporate the perturbation of the compactified ${\hat c}=1$
theory by both momentum and winding modes. The same problem arises
in the equivalence of the topological B-model on the deformed conifold and
$c=1$ bosonic strings at the self-dual radius.

Since the compactified ${\hat c}=1$ 0A theory is T-dual to the ${\hat c}=1$ 0B
theory on the dual radius, our results suggest that the topological B-model on that geometry is
equivalent to ${\hat c}=1$ 0B string theory at the radius
$R=\sqrt{2\alpha'}$. It would be interesting to prove this directly by finding different
fermionic realization of the topological B-model.

In this paper we found the equivalence of two theories only at the
perturbative level. This is natural as the topological string theory
can be defined only perturbatively. One may consider the ${\hat
c}=1$ theory at the radius $R=\sqrt{\alpha'/2}$ as the
nonperturbative completion of the topological B-model on the
deformed $\Z_2$ orbifolded conifold. For the recent discussion on
the nonperturbative approach to topological strings,
see~\cite{Park:2000au,Hofman:2002rv,Nekrasov:2004js,Kapustin:2004jm,Dijkgraaf:2004te,Nekrasov:2004vv}

\section*{Acknowledgments}
We would like to thank Jae-Suk Park for valuable
discussions. S.H., K.O. and J.-D. P. would like to thank the Korea
Institute for Advanced Study (KIAS) for its hospitality, where
this work has been completed. This work was supported by
the Korea Research Foundation Grant KRF-2004-042-C00023.



\end{document}